\documentstyle[12pt]{article}
\begin{document}

\title{Charge frustration in complex fluids and in electronic systems}
\date{\today}
\author{Carlo Carraro\\
Department of Chemistry, University of California\\
 Berkeley, California 94720-1460  USA}
\maketitle
\begin{abstract}
The idea of charge frustration is applied to describe the properties of
such diverse physical systems as oil-water-surfactant mixtures and
metal-ammonia solutions.
The minimalist charge-frustrated model possesses one energy scale and two
length scales. For oil-water-surfactant mixtures, these parameters have
been determined
starting from the microscopic properties of the physical systems under
study. Thus microscopic properties are successfully related to the observed
mesoscopic structure.
\end{abstract}
\newpage
\section{Introduction}

The physics of frustration underlies the behavior of many different
physical systems.
Competing interactions or unfavorable boundary conditions often frustrate
the phase equilibrium of a physical system and play an important role in
the formation of large interfacial structures or of topological defects.
Examples include the domain walls in a magnetic system or in
oil-water-amphiphile mixtures, the vortex lines in a type-II
superconductor, or the entangled disclination lines in a metallic glass. At
sufficiently low temperature, these defects can exist in an ordered state,
such as the hexagonal vortex arrays of type-II superconductors, or the
lamellar, hexagonal, and cubic phases of an amphiphilic mixture. At higher
temperature, these ordered structures melt into disordered (liquid) phases,
such as the vortex liquid phase of high temperature superconductors or the
microemulsion phase of amphiphilic mixtures.

These disordered liquid phases behave differently than simple liquids. They
retain a great deal of organization on mesoscopic scales, as is beautifully
exemplified by small angle neutron scattering experiments on
water-oil-amphiphile mixtures. The presence of order on scales that are
much larger than molecular dimensions makes it difficult to explain
experimental observations starting from the known molecular properties of
the system.
Recent progress in this direction has been possible using the idea of
charge frustration\cite{wcc,wcc2}. 

Charge frustration originates in the impossibility to separate opposite
charges on macroscopic scales, due to the overextensive energy cost that
such separation would entail. Thus, in a neutral Coulomb gas, charge
neutrality is a local effect, and interactions have a finite range,
$\Delta$, of the order of the Debye-H\"uckel screening length. 

Consider, for example,
a two-component system in which short range forces favor phase separation.
If the two components carry opposite charge, macroscopic phase separation
cannot take place. Nevertheless, it is possible that the length scale over
which charge neutrality is enforced, $\Delta$, is much larger than the
range, $a$, of the forces which drive phase separation. In this case, the
two-component system can find local rearrangements that minimize the short
range repulsion while keeping overall neutrality on a length scale
$\Delta$. The compromise can be achieved, for example, by forming narrow
interfaces separating mesoscopic domains. This arrangement is called
microphase separation. The material is homogeneous on large macroscopic
scales, but phase separated on shorter mesoscopic scales. 

In this paper, we study charge frustration in microemulsions and in some
electronic systems. In microemulsions, which we discuss in Sec. II, charge
frustration does not originate from the presence of unbound electric
charge; rather, its origin is in the entropy reduction that stoichiometric
constraints impose on an amphiphilic mixture. In Sec. III, we discuss
metal-ammonia solutions, an electronic system where charge frustration is
literally brought about by electric charge. Our conclusions are presented
in Sec. IV. 

\section{Charge frustration in oil-water-amphiphile mixtures} 

Water and oil are immiscible fluids: when mixed, they phase separate into
two almost pure components, divided by a narrow interface. The magnitude of
the interfacial tension, of order 20 dynes/cm, is a measure of the strength
of the hydrophobic interaction that drives the phenomenon of phase
separation; the width of the interfacial region, of order the size of a
water molecule, is a measure of its range. Thus, the phase diagram as
well as the long wave length (Ornstein Zernicke) structure of the binary
mixture can be described simply in terms of one characteristic energy
parameter and one length scale. A detailed molecular description is indeed
redundant and conveniently forgone in favor of a density field theory, with
hamiltonian given by
\begin{equation}
{H\over k_BT}=\int\!\! d^3r \biggl(\!-\rho_w(\vec r)\log\rho_w(\vec r)- 
{\rho_o(\vec r)\over n_o}\log\rho_o(\vec r)+\!\!\!\sum_{ab\in
(w,o)}\!\!\!\rho_a(\vec r){J_{ab}\over k_BT}
(r-r')\rho_b(\vec r')\biggr), \label{eq:density}
\end{equation}
where $\rho_w$, $\rho_o$ represent water and oil densities, respectively,
and the interaction matrix $J$ is characterized by a short range, $a$. An
equally convenient representation can be given in terms of a spin 1/2 Ising
model:
\begin{equation}
H=-\sum_{ij}J_{ij}s_is_j-h\sum_{i}s_i.
\label{eq:ising}
\end{equation}
Here, up (down) spins represent water (oil) density. 

The addition of a small amount of surfactant molecules frustrates phase
separation. This is because the polar head of a surfactant interacts more
favourably with the polar water molecules, while the hydrophobic tail of a
surfactant is driven towards the oil phase. The presence of a covalent bond
between head and tail of a surfactant implies that their amphiphilic
tendency is satisfied if both water and oil phases can be found within a
spatial range roughly equal to the length of a surfactant molecule. Thus,
surfactants promote the formation of oil-water interfaces. 

The presence of
extensive interface (i.e., of amounts of interfacial area proportional to
the number of surfactants) implies that an imaginary straight line through
the amphiphilic mixture must thread many oil-water interfaces, a fact which
is incompatible with macroscopic phase separation. Indeed, the mixture
consists of a multitude of oil and water domains. It is microphase
separated.

Small angle scattering experiments\cite{sstrey} show that the system is
indeed ordered
on mesoscopic scales, in the sense that there is a well defined domain
size, which leads to oscillations in the long wave length structure factor.
It is a challenge for theory to predict this domain size, starting from the
knowledge of the chemical composition of the mixture. 

Amphiphilicity is a molecular property, but phase separation is best
described in terms of density fields. To successfully develop a theory of
microphase separation, one must address the question: how do density fields
"know" about molecules?
To answer this question in a rigorous manner, one should a) define
appropriate density fields, and b) integrate out the molecular coordinates
in favor of the collective variables.
Neither step is trivial.

An imaginative proposal for an approximate solution to this problem was put
forward by Stillinger\cite{stil}, who chose as collective variables the
densities of hydrophobic and polar species; i.e., the natural variables in
term of which the problem of phase separation is simple. However, this
choice entails breaking up amphiphilic molecules into polar heads and
hydrophobic tails. Stillinger further breaks up each head (tail) into $n_H$
($n_T$) monomers, and assigns a polar density field, $\rho_H$, to the head
monomers and a hydrophobic density, $\rho_T$, to the tail monomers.
Disregarding the molecular nature of the mixture, the Hamiltonian would be
as in Eq.~(2.1). However,
the interaction between density fields must reflect (at least on a mean
field level) the covalent bond present in the molecules. Therefore,
assuming that fluctuations in the head-tail mixture can still be described
by gaussian fields, one can write \begin{equation}
{H\over k_BT}={1\over 2}\int d^3r\int d^3r' \sum_{ab}\rho_a(\vec r)\Omega_{ab}(r-r')\rho_b(\vec r'), \ \ \ \ \ (a,b)\in(H,T),
\label{eq:st1}
\end{equation}
but the interaction matrix $\Omega_{ab}$ remains to be determined. 

The appeal of Stillinger's work lies in the way it makes the connection
between molecular variables and density fields, by comparing the
intramolecular structure factor, which can be computed independently, in
the low density approximation, starting from either molecules or densities.
In terms of densities, one has
\begin{equation}
\Omega^{-1}_{ab}(\vec k)=<\rho_a(\vec k)\rho_b(\vec -k)>. \end{equation}
In terms of molecular coordinates, one finds \begin{equation}
<\rho_a(\vec k)\rho_b(\vec -k)>=\sum_{i\in a}\sum_{j\in b} <e^{i\vec
k\cdot\vec r_i}e^{-i\vec k\cdot\vec r_j}>\approx n_an_b-{k^2\over
6}\Delta^{(2)}_{ab},
\end{equation}
where
\begin{equation}
\Delta^{(2)}_{ab}=\sum_{i\in a}\sum_{j\in b}<|\vec r_i-\vec r_j|^2>.
\end{equation}
This comparison enables
one to identify the interaction matrix, which, along with the hydrophobic
interaction, contains a Coulomb-like part: \begin{equation}
\Omega_{HT}(\vec k)\approx{3\over n_Hn_T \Delta^{(2)}k^2}, \end{equation}
where
\begin{equation}
\Delta^{(2)}={\Delta^{(2)}_{HT}\over n_Hn_T}- {\Delta^{(2)}_{TT}\over
n_T^2}-{\Delta^{(2)}_{HH}\over n_H^2} \end{equation}
is of order the mean square length of a surfactant molecule. A mapping of
an amphiphilic mixture onto an electrolytic solution naturally ensues. Note
that the effect of the Coulomb interaction is such that polar and
hydrophobic densities attract, thus opposing phase separation. Hence the
origin of charge frustration.

In spite of the apparent long range nature of the interaction matrix, the
correlations in the system are in fact short ranged, as they must be in any
fluid away from a critical point. The equivalent electrolytic solution is
neutral, and the Coulomb potential is thus screened. An ion in a neutral
Coulomb gas is surrounded by an oppositely charged cloud of ions within a
Debye screening length. Similarly, in an amphiphilic mixture, the presence
of surfactant head (tail) density at a point implies the presence of an
equal amount of tail (head) density within a distance of order the length
of a surfactant molecule. 

This observation is crucial in determining the way in which the elementary
charge should be assigned in the electrolyte-amphiphile mapping. Indeed,
note that a linear response argument does not uniquely identify the
magnitude of an elementary charge. (One encounters examples where
frustration expresses itself through the presence of charges that are
unambiguously "quantized," e.g., when topological constraints are present,
or when charge frustration is quite literally due to the presence of
electric charge, as in Sec. III below.)

Stillinger's original method\cite{stil}
divides up each polar head and hydrophobic tail into monomers, and assigns
a unit of charge to each monomer. Suppose for simplicity that $n_H=n_T$.
Then, putting $\beta=1/k_BT$, the unit charge is given by \begin{equation}
q=\left(\frac{3}{4\pi\beta\rho\Delta^{(2)}n^2}\right)^{1/2}, \end{equation}
where $\rho$ is the number density of amphiphiles. This assignment leads to
a Debye length
\begin{equation}
\lambda= \biggl({n\Delta^{(2)}\over 6}\biggr)^{1/2}\propto n^{3/2}.
\end{equation}
This result is obviously
unphysical, since it violates stoichiometry on length scales that can be as
large as many surfactant molecule lengths. This contradiction has gone
unnoticed in some subsequent work\cite{deem}. 

It is easy to understand why Stillinger's assignment leads to incorrect
results for long surfactant chains. The repulsion of equally charged
monomers on a surfactant head (or tail) leads to the unphysical "swelling"
of the molecule. This is avoided by lumping the charges into a single unit
per head (tail) of surfactant molecule. This alternative procedure, first
introduced by Wu, Chandler, and Smit\cite{wu}, leads to a Debye
length that scales $linearly$ with the number of monomers in a surfactant
molecule, which is the correct result. A convenient representation of
charge frustration of the Ising model ensues: additional site variables,
$t_i=0,1$, are introduced to tag a fraction of the spins. To each tagged
spin $(t_i=1)$ one assigns a charge, which is positive if the spin is up,
and negative if the spin is down.
The charge frustrated Ising Hamiltonian can thus be written as \begin{equation}
H=-\sum_{ij}J_{ij}s_is_j-h\sum_{i}s_i+
Q^2\sum_{ij}{s_it_is_jt_j\over |r_i-r_j|}+\mu\sum_{i}t_i, \label{eq:wcs}
\end{equation}
with the correct value of the charge being\cite{wcc} \begin{equation}
Q=\left(\frac{3}{4\pi\beta\rho\Delta^{(2)}}\right)^{1/2}. \end{equation}
The number of tagged spins, i.e., the density of amphiphiles $\rho$, is
controlled by the chemical potential $\mu$. 

Tracing out the variables $\{t_i\}$ leads to a reduced spin 1/2
Hamiltonian, whose statistical mechanics has been studied in detail
elsewhere\cite{wcc}. Figure~1 is an indication of the agreement between the
measured structure factor and calculations based on the charge frustrated
model. Our previous work has shown that the reduced spin 1/2 Hamiltonian,
while able to predict the structure of microemulsions with remarkable
accuracy, cannot accurately predict their phase diagram; more specifically,
it does not predict three-phase coexistence. This deficiency is not at all
surprising, since tracing out the variables $\{t_i\}$ amounts to averaging
over the density of surfactant molecules. 

To remedy this shortcoming,
it is convenient to perform, on each site, the change of variables
\begin{equation}
s=\xi+\eta(1-\xi^2); \ \ \ \ t=1-\xi^2
\end{equation}
where the new variable $\xi=0,\pm1$ describes density fluctuations in the
ternary mixture, and $\eta=\pm1$ describes charge fluctuations. Effecting
this transformation yields the Hamiltonian of a charge-frustrated spin 1
Ising model. There are two main advantages of this transformation over the
procedure followed in Ref.\cite{wcc}. First, surfactant density
fluctuations are decoupled from charge fluctuations, allowing the latter to
be integrated out easily, at the Gaussian (Debye-Huckel) level. Second, a
proper treatment can be given of hydrogen bonding effects, which cause the
water-head interaction to be different from the oil-tail interaction, and
are principally responsible for the characteristic shape of the three-phase
coexistence region in the phase diagram of oil-water-amphiphile ternary
mixture\cite{gompschick}.

\section{Charge frustration in metal-ammonia solutions} Alkali metals
dissolve readily in liquid ammonia, and the resulting solution has been
known for over a century to possess remarkable electronic and optical
properties\cite{old}. They can all be traced to the favorable solvation
energy of an alkali ion, which is larger in magnitude than the ionization
energy of an alkali atom. Thus, metal-ammonia solutions consist of unbound
electrons immersed in a sea of ammonia molecules and solvated alkali ions.
As the metal concentration rises from zero, these solutions are observed to
behave at first essentially as insulators (except for ionic conductivity),
until a concentration is reached, where the larger electric conductivity
signals the onset of extended electronic states. This behavior can be
interpreted as a delocalization effect. However, there are reasons to
suspect that the conducting phase possesses a nontrivial structure, as is
hinted to by the recent quantum molecular dynamics simulations of
Ref.~\cite{klein}.
The arguments presented below point to charge frustration and to the
resulting possibility of microphase separation in this system. 

Ammonia is an electron-rich molecule. The Pauli principle implies that an
electron immersed in ammonia feels a strongly repulsive potential when it
resides on or near an ammonia molecule. Let us neglect for the moment the
existence of the Coulomb interaction, i.e., let us turn off the charge of
the electrons. Then, the only way that electrons can mix with ammonia is by
taking advantage of density fluctuations in the liquid, which generate
small empty cavities, in which an electron may be localized. Cavities that
are large enough to host electrons, without excessive cost in zero point
energy, are rare. Thus, at a finite electron concentration, confinement
becomes unfavorable, and the electronic states become extended. A competing
process can take place, however, when the Fermi level reaches the
difference in chemical potential between liquid vapor (see Fig.~2). Since
the cost of adding electrons in the vapor is essentially zero, the system
phase separates into two coexisting phases: electrons in liquid ammonia and
electrons in vapor. One can schematically regard the system as an ideal
fermion gas,
phase separated from liquid ammonia. The interfacial tension is given by
the surface tension of ammonia plus a contribution from the inhomogeneous
electronic density near the surface of the liquid. Phase separation of
ideal fermions from a sea of classical "blockers" has been observed in the
computer simulations of Alavi and Frenkel\cite{ala}. 

The picture described above neglects the Coulomb interaction between
electrons and the solvated alkali ions. As we discussed in the previous
section, oppositely charged densities cannot phase separate over
macroscopic length scales. Phase separation must therefore be charge
frustrated, and we may expect that some of the conclusions we
arrived at in our study of microemulsions hold true for metal-ammonia
solutions as well. Note that in this system charge frustration is not an
entropic effect; it is brought about by the interaction potential. The
parameters involved are thus unambiguously identified. An interesting new
feature is the quantum mechanical nature of one of the components. While a
quantitative treatment of this
additional feature is left for future work, we outline here the qualitative
behavior of the system. 

We note that metal ammonia solutions should be isomorphic to binary
(water-amphiphile) mixtures, ammonia playing the role of water and the
electron and solvated ion densities playing the role of surfactant tail and
head densities, respectively. One must identify the basic energy and length
scales of this problem. We have mentioned already interfacial tension and
width of the "bare" (charge-free) system, whose role is completely
analogous to that played in the binary mixture counterpart.

We expect that an additional length scale will be important, which
characterizes the range of the frustrating interaction. To estimate it, let
us consider an interface in the liquid. The appropriate energy diagram,
shown in Fig.~3, is obtained
from Fig.~2, with the addition of the Coulomb interaction. We see that the
Coulomb interaction allows for the presence of electronic bound states near
the surface of the liquid. The energy gained from filling up those states
will compete with the bare liquid vapor interfacial tension, to produce a
low tension interface, which can become extensive.

We expect electron-ammonia bicontinuity to result as a consequence. This
expectation is reinforced by the observation that the frustrating length
scale, i.e., the length over which the system restores charge neutrality,
should be comparable to the length over which the electron bound states
extend outside the surface of the liquid. Given the large dielectric
constant of liquid ammonia, this length will be well in excess of a Bohr
radius; in fact, it could be several times larger than the surface width of
the surface of the liquid. A precise estimate involves computation based on
density functional theory, which we leave for future work. Based on our
work on amphiphilic systems, however, we predict that a screening length of
order ten Bohr radii will be sufficient for bicontinuity to produce
observable effects in the small angle structure factor of metal-ammonia
solutions. Indeed, this length corresponds to the mean square length of
short polyglycol ether surfactants. Unfortunately, at present, experimental
information on the long wave length structure of these solutions is not
available to us.

\section{Conclusion}
Frustration of phase separation by a long range attractive interaction
(such as the Coulomb potential) has
spectacular consequences on the structure and phase diagram of many
physical systems. We have discussed in detail amphiphilic mixtures (which
include diblock copolymers as an interesting limit) and metal-ammonia
solutions. A simple unifying description of such diverse systems is
provided by charge frustrated Ising models, such as the spin 1 model
defined by the Hamiltonian (\ref{eq:wcs}). 

Because Ising models are employed in so
many and so diverse areas of physics, one could speculate that other
frustrated physical systems exist, which can be described by
charge-frustrated Ising models, and, thus, exhibit the rich structural and
phase behavior we have discussed in the previous sections. For example,
will the presence of vacancies in magnetic systems cause (micro)phase
separation? The question of phase separation has been raised recently in
relation to the behavior of cuprate superconductors\cite{kiv1}, such as
La${}_{2-x}$Sr${}_x$CuO${}_4$. The undoped material is an antiferromagnetic
insulator, while a high enough concentration of vacancies (typically, a few
percent) destroys antiferromagnetic order, and produces a poorly conducting
metal, which becomes superconducting at low temperature. Semiclassical
expansions of the two dimensional $t-J$ model \cite{sorella} have produced
evidence
for phase separation of the vacancies from the spins in these materials.
Vacancies, however, carry charge, which gives rise to long range repulsion.
This repulsion, which is ignored in the $t-J$ model, would prevent
macroscopic phase separation, but could lead to microphase separation.

The example above illustrates an interesting novel feature of charge
frustration, namely, that frustration is induced by the presence of charges
of the same sign. In the language of charged fluids, one deals here with
one component plasmas, rather than with the neutral Coulomb gas we
encountered in Sec.~II. Nevertheless, screening can be expected again to
play an important role, by bringing about a second length scale in addition
to the short (nearest neighbor) range of the exchange interactions. 

Finally, we mention the possibility
that a similar one-component plasma mapping might be established, following
the work of Kivelson $et\ al.$, between charge-frustrated models and models
of structural glasses and supercooled liquids\cite{kiv2}. The details of
this mapping, as well as the origin and magnitude of the additional length
scales or of the elementary ``charges,'' remain to be discovered.

I am indebted to David Chandler and Hyung-June Woo for many stimulating
discussions. This work was supported by the National Science Foundation
under grant CHE 9508336 and by the Office of Naval Research under grant
N00014-92-J-1361.

\newpage
\noindent{ \bf 
FIGURE CAPTIONS	}

\flushleft{1:}
 Small angle structure factor of AOT--D${}_2$O--$n$-decane mixtures, at
varying volume fraction $f$.
(a) $f=0.181$; (b) $f=0.237$; (c) $f=0.323$. The symbols represent the
experimental data from
\protect\cite{sk} The solid lines are the results of calculations based on
the charge frustrated
model. We have expanded the inverse long wave length structure factor in
powers of the momentum
transfer, and truncated the expansion at the fourth order, to make
connection with the Teubner-Strey
formula \protect\cite{teub}

\flushleft{2:}
 Energy level diagram for ideal fermions in a liquid. Fermions with energy
below $E_c$ are
localized in cavities inside the liquid. The chemical potential difference
between vapor and liquid,
$\mu_v-\mu_l$, represents the energy cost of forming an infinite cavity in
the liquid.

\flushleft{3:}
 Energy level diagram for electrons in metal-ammonia solution. This diagram
is obtained from
Fig.~2 by ``switching on'' the charges of the electrons and of the solvated
ions. Note the discrete
Coulomb-like spectrum of electronic states near the surface of the liquid.
The electronic states
inside the liquid are different from Fig.~2, due to the strong polar nature
of the solvent.

\end{document}